\documentstyle[preprint,aps]{revtex}

\begin{document}
\draft
\title{\bf    Global  solutions of a free boundary problem for 
selfgravitating scalar fields.}
\author{Edward Malec}
 \address{  Institute of Physics, Jagellonian University,
30-059 Cracow, Reymonta 4, Poland}

\maketitle
\begin{abstract}
The weak cosmic censorship hypothesis can be understood as  a statement that
there exists a global Cauchy evolution of a selfgravitating system
outside an event horizon. The resulting Cauchy problem has a free
null-like inner boundary. We study a selfgravitating spherically symmetric 
nonlinear scalar field. We show  the global existence of a spacetime with a 
null inner boundary that initially is located outside the Schwarzschild 
radius or, more generally, outside an apparent horizon. 
The global existence of a patch of a spacetime that is exterior to an event  
horizon  is obtained as a limiting case. 
\end{abstract}
\vskip 2cm
 
\centerline{1. Introduction}

The cosmic censorship hypothesis  \cite{Penrose} can be informally
stated as {\it singularities are hidden inside black holes}.
Various attempts to formalize that statement have led to a collection
of results that can be assembled into two categories, one of which
may be called "geometric"  and the other "dynamic".

The first approach, reviewed for instance in \cite{Tipleratal} and
\cite{Wald}, seems to be strongly influenced by the post-Leray
\cite{Leray} notion of the  global hyperbolicity.  The latter requires
(in addition to the standard hyperbolicity condition as stated for instance
in  \cite{Pietrovski}) stringent smoothness properties of
coefficients of a hyperbolic operator as well as  a causality condition. In
a related stream of research a considerable   effort is put into the
examination of geometric quantities in various models,  in order to define
a notion of a singularity and then, to identify those singularities that
must be enclosed by event horizons. A singularity can be understood as a
point of space-time having divergent curvatures; a cosmic censorship
hypothesis would require that all geodesics originating at a  singularity
remain inside a black hole. The post-Leray apparatus of ideas makes  it
plausible, however, to  associate singularities  with the existence of
geodesics having a finite length \cite{Penrose 1965}.
On the other hand, an increasing number of counterexamples to
resulting  versions of the cosmic censorship hypothesis has given
inspiration to various notions of truly "singular" singularities, of which
I mention a subclass of the so-called strong singularites \cite{Krolak}
that is associated with the existence of apparent horizons
\cite{Penrose 1965}.

 The dynamic approach has been initiated, to my knowledge, by Eardley
 and Moncrief \cite{Moncrief}. It bases on the notion of a Cauchy
 problem as formulated, for instance,  in \cite{Pietrovski}.
 A weak version of the cosmic censorship hypothesis can be formalized as
 follows, in the case of asymptotically flat spacetimes:

 {\it Given "reasonable"  asymptotically flat initial data of
 Einstein-matter field equations and assuming "reasonable" energy
 conditions, there exists a Cauchy evolution that is global in
 the sense, that a solution does exist outside  black holes for arbitrarily
 large values of the time of an asymptotic observer .}

It is possible to formulate a " dynamic" version of the cosmic censorship in
cosmological models\cite{Moncrief}, where the geometric approach runs
into a trouble  because there exist  serious conceptual problems in defining
a notion of a black hole. Most of the existing literature is concerned with
the validity of that version of the cosmic censorship in
various cosmological models \cite{Chrusciel}.

An analysis of spherically symmetric systems suggests that
the two approaches to the cosmic censorship  can converge.
Trapped surfaces, that are inherent to the study of the cosmic censorship
 conjecture  by Kr\'olak \cite{Krolak}, have been investigated in spherically
symmetric geometries
\cite{BMOM1988}, \cite{MOM1994b}. Their appearance is always connected with
a large concentration of matter. If a  sphere $S$ that is centered around
a point of symmetry is trapped, then  it contains amount of matter
$M(V)=\int_{V(S)}\rho dV$ of the order $L(S)$ \cite{MOM1994b}, where $V(S)$
is the volume inside $S$ and $L(S)$ is the geodesic radius of $S$.
And conversely, a large amount of matter, $M(V)>L(V)$, leads to the
formation of an apparent horizon.
The quantity $\int_{V(S)}\rho dV$  can be bounded from above and from below
by a suitable Sobolev norm (usually $H_1$)  of a matter field. Sobolev norms
are, in turn, natural objects in the dynamic approach; an evolving system
does exist so long as its Sobolev norm is finite. Singularities can be
defined as those regions of space-time that give an infinite contribution
to a Sobolev norm. Take an initial configuration
of a compact support of a hyperbolic  system of matter fields coupled to
Einstein equations. Hyperbolicity (I always use the Petrovsky meaning
of that term; see Section 3) means that the support of matter
remains finite during a finite evolution. Assume that a Cauchy evolution
can be analyzed in the Sobolev class $H_1$. If an evolution
breaks down then there exists a sphere $S$ such that a contribution to  the
Sobolev norm $H_1$ coming from the interior of $S$ becomes big in comparison
to $L(S)$ - hence there must exist a trapped surface and an event horizon
enclosing $S$.  That means that, in the spherically symmetric case at
least (and under the assumption that there exists an evolution in $H_1$ -
that fact is not obvious), the  notion of   strong singularity is related
to the notion of dynamic singularity.

The "dynamic" version of the weak cosmic censorship hypothesis leads  to
an  external Cauchy problem with a free inner boundary being a null cone,
as is explained below.
 Take a four-dimensional (asymptotically flat)
Lorentzian manifold $M$ and define its space-like foliation by hypersurfaces
$\Sigma_t$, where $t$ is the asymptotic time, $t\le \infty $. A black hole
$H$ can be defined as the largest piece
of $M$ that still can be enclosed by a null cone $\delta H$ such that
the area of  the
intersection  $\delta H \cap  \Sigma_t$   remains uniformly bounded.
(That definition is motivated by the fact that  an area of a  spherical black
hole is bounded from above  by $16\pi m^2$, where
$m$ is the asymptotic mass \cite{MOM1994a}.) The weak cosmic censorship
hypothesis reads in this context as follows: for "reasonable" initial data
and "reasonable" matter, there should exist a global  Cauchy evolution of
a region exterior to $H$. The inner boundary (event horizon)
$\delta H$ of that region is a null cone so
that we arrive naturally to an external Cauchy problem with an event horizon
as a free  null-like inner boundary.

In this paper I will study a more general case of a Cauchy evolution with
a free null-like boundary that does not necessarily coincides with an event
horizon. A selfgravitating nonlinear scalar field is used as a matter model.
The main results of the paper are contained in Theorems 1 and 2 of Section 4.
Their formulation would require a number of preliminary definitions,
so instead let me just discuss the most important points.

The fact that I find the most interesting is that even the local version
of the Cauchy problem requires the positivity  of the potential
(selfinteraction)   term $W(\phi )\ge 0$.
Using a technical jargon, negative selfinteraction
$W(\phi )$  means  that scalar field equations (see
\ref{3.6}, \ref{3.7}) would loose their strict hyperbolicity. (\ref{3.1})
implies   that the loss of hyperbolicity is particularly plausible inside
a region enclosed by the Schwarzschild radius $2m$.

The  smoothness of  an evolving solution is locally preserved, that is its
differentiability properties are kept intact for a small enough interval of
time. The global existence is also  proven, through standard apriori
estimates, in the  differentiability class $\phi \epsilon H_k$, ($k\ge 3$).
 Having the global existence outside a null cone $\delta H$, one is given
a geometry of $\delta H$ and a scalar field on $\delta H$; they in
turn can be regarded  as initial values of the characteristic Cauchy problem.
The present  work is supplementary to the investigation of
Christodoulou \cite{Christodoulou}, who  investigated the existence of a
solution of the characteristic Cauchy problem for (massless) scalar fields.

Although I deal only with scalar fields, my impression is that
the formalism presented below is capable to include other forms of
nontachyonic matter, that is
assuming that matter fields  are described by hyperbolic equations
and their energy-momentum tensor satisfies some energy conditions.
Spherically symmetric Yang - Mills and SU(2) nonlinear sigma model do in
fact satisfy the needed conditions.   
Einstein - Vlasov system  does not belong to that category  that
can be analyzed with techniques presented here, but let me mention that
a recent work  of Rein, Rendall and Schaeffer shows   that there exists a
 global solution  for Einstein - Vlasov system \cite{Alan9295},
 if matter does not form
a singularity at the symmetry center.

The content of the rest of the paper is following. Section 2 presents
the Einstein equations coupled to a spherically symmetric field equations
in (1+3) splitting. They can be
reduced to a system of integro-differential equations with  matter
fields being  dynamical variables. The reader may find it
useful that  one can explicitly express components of the  spherically
symmetric metric as some functionals of matter-related terms - see formulae
(\ref{2.14} - \ref{2.18}) - in any splitting of the space-time, that
is for any value of the trace $trK$ of the extrinsic curvature.

Section  3 describes the Einstein - scalar field equations in the so-called
polar gauge. That way of foliating  a space-time is particularly
convenient when discussing the external Cauchy problem with a free inner
boundary.  In most gauges one is confronted with the need to impose
an additional boundary condition at the inner boundary; in contrast with that
the polar gauge does not require any extra conditions.

Section 4 contains the main result, Theorems 1 and 2. 
The proof of the local Cauchy solvability bases on the standard
compactness method. The main part of the  local part of the
proof of Theorem 1 is relegated  to Section 5 while the global 
part is based on important $L_p$ and $H_l$ 
global  estimates that are proven in Section 6.  In Section 6 a change of
dependent variables leads to a set of first order 
equations that are "almost" linear; that  allows one to obtain a set
of apriori global estimates.
Section 7 discusses a 
generalization of the proof of Theorem 1 that finally leads to the proof of 
Theorem 2.

\vskip 2cm
\centerline{ 2. Equations.}

  The most general metric of a spherically symmetric spacetime
is given by the expression
 \begin{equation}
ds^2 = -  N^2(r, t)dt^2 + a(r, t) dr^2 +  R^2(r, t)  d\Omega^2,
\label{2.1}
\end{equation}
where $t$ is a time coordinate, $r$ is a radial coordinate, $R $
is the areal radius and $d\Omega^2 = d\theta^2 + \sin^2\theta d\phi^2$ is the standard line
element on the unit sphere with with the angle variables $0\le \phi < 2\pi $
and $0\le \theta \le \pi $.  I assume that at spatial infinity
  $N=1$ (hence $t$ coincides with the proper time of an observer who
  is far away from the center of symmetry) and $a=1$.

    Initial data of Einstein-scalar fields equations on an exterior part
    $\Sigma_t^{out}$ of a     Cauchy slice $\Sigma_t $
consist of   $(g_{ij}, K_{ij}, \rho, j_i)$ where $g_{ij}$ is the intrinsic
metric, $K_{ij}={\partial_tg_{ij}\over 2N }$ is the extrinsic
curvature, $\rho =-T_0^0 $ is the matter energy density and $j_i=NT_i^0$
is the matter current density. $T{\mu \nu }$ is the energy - momentum tensor.
Initial data  satisfy the constraints
 \begin{equation}
^{(3)}R-K_{ij}K^{ij}+(trK)^2=16\pi \rho
 \label{2.2}
\end{equation}
 \begin{equation}
 \nabla_iK^{ij}-\nabla^jtrK =-8\pi j^j
 \label{2.3}
\end{equation}
where $^{(3)}R$ is the scalar curvature of the intrinsic metric.
Above (and elsewhere in the paper) I assume the Einstein summation
rule with the exception of indices $r, \theta $ and $ \phi $ whose
repetition  is supposed never to mean summation. Latin indices
change from 1 to 3 while Greek indices range from 0 to 4.

In the spherically symmetric case it is  convenient to formulate
the whole set of  Einstein equations in terms of the extrinsic curvature
$K_{ij}$ (which describes the embedding of the three-dimensional hypersurface
$\Sigma $ into a spacetime) and the mean curvature $p$ (which is
a trace of the two-dimensional extrinsic curvature that describes
the embedding of a two-dimensional   sphere centered around
the symmetry center into $\Sigma_t $).
The interior metric of  $\Sigma_t $ is
 \begin{equation}
ds^2_{(3)} =   a  dr^2 +  R^2  d\Omega^2.
\label{2.4}
\end{equation}
  The  mean curvature of a centered two-sphere as embedded in
an initial three dimensional hypersuface  is
\begin{equation}
p= {2\partial_rR\over  \sqrt{a} R}.
\label{2.5}
\end{equation}

If the trace $trK$ of $K_{ij}$ is fixed then there is only
one independent  component of the   extrinsic curvature, say the radial-radial
component $K=K_r^r$,
and the remaining two components are each equal to ${trK-K\over 2}$,
 \begin{equation}
trK-K =2 K_{\theta }^{\theta } =2 K_{\phi }^{\phi }.
\label{2.6}
\end{equation}

In terms of  $K$ and $p$ the constraints can be written as
\begin{eqnarray}
{\partial_r (  p R )\over \sqrt{a}}= -8\pi R\rho  -{3R\over 4}(K)^2+{R\over 4}
(trK)^2+{R\over 2}KtrK -{Rp^2\over 4} +{1\over R}
\label{2.7}
\end{eqnarray}
and
\begin{eqnarray}
{\partial_r ( R^3( K -trK) )\over \sqrt{a}}= -8\pi R^3 {j_r\over \sqrt{a}}
-ptrKR^3
\label{2.8}
\end{eqnarray}
In the spherically symmetric case the full set of the Einstein equations
consists of the  two preceding ones,  the evolution equation
 \begin{eqnarray}
 \partial_0(  K -trK )= {3 N \over 2}(  K)^2 +{N\over 2}(trK)^2-2NKtrK
 -{ p^2R\over \sqrt{a}}\partial_r { N \over pR}+8\pi N(T_r^r +\rho )
 \label{2.9}
 \end{eqnarray}
and  the lapse equation
 \begin{eqnarray}
\nabla_i \partial^i  N  = N \Bigl( {3  \over 2}(  K)^2  + (trK)^2/2
- KtrK+ 4\pi  (\rho + T_i^i)    \Bigr) +\partial_0trK .
 \label{2.10}
\end{eqnarray}
The Einstein equations and Bianchi identities yield the energy-momentum
conservation equations,  which in the case of spherical symmetry reduce to
 \begin{eqnarray}
 \partial_0{j_r\over \sqrt{a}} +N(K+trK ){j_r\over \sqrt{a}}+
 {N\over \sqrt{a}}\partial_r T_r^r
 +{\partial_rN\over \sqrt{a}}(T_r^r+\rho)+Np(T_r^r-T_{\phi }^{\phi })=0
 \label{2.11}
\end{eqnarray}
 \begin{eqnarray}
- \partial_0\rho -{N\over \sqrt{a}} j_r-{N\over \sqrt{a}}\partial_r {j_r\over
\sqrt{a}} -{2\partial_rN\over a}j_r-NK(T_r^r-T_{\phi }^{\phi })
-NtrK(\rho +T_{\phi }^{\phi })=0.
 \label{2.12}
\end{eqnarray}
Using the above equations, one can express metric coefficients in terms
of $T_{\mu \nu }$ and $trK$. From the momentum constraint one finds
\begin{equation}
RK(R)-RtrK(R)={C+8\pi \int^{\infty }_Rj_rr^3dr\over R^2}+2{\int_R^{\infty }
trKr^2dr\over R^2}.
\label{2.13}
\end{equation}
The parameter $C$ is constant on a particular Cauchy slice and it must be
set to 0 on slices including the symmetry center.  $C$ is arbitrary, however,
on slicings that do not include the world line $R=0$.

The integration of the hamiltonian constraints leads, after some
algebra, to the expression
\begin{equation}
pR=2\sqrt{1-{2m\over R}+{2m(R)\over R}+{R^2\over 4}(trK-K)^2},
\label{2.14}
\end{equation}
where $m$ is the asymptotic (ADM) mass. The function $m(R)$  can be
interpreted as a local energy energy density (it is easy to notice that
$m(0)=m$) and it is given by the equation
\begin{equation}
m(R)= 4\pi \int_R^{\infty }dr r^2(\rho +rj_r(trK-K)).
\label{2.15}
\end{equation}
The lapse $N$ can be determined from (\ref{2.9}),
\begin{equation}
N={pR\over 2}\Bigl( 1+4\int_R^{\infty }{\partial_t(Kr^3-trKr^3)dr
\over \beta (r)p^3r^5}\Bigr) \beta (R)
\label{2.16}
\end{equation}
where
\begin{equation}
\beta (r) = e^{  \int_r^{\infty }(16\pi (-T_r^r-\rho )-2K trK +2(trK)^2)
{1\over p^2s}ds}.
\label{2.17}
\end{equation}
The lapse $N$ satisfies (\ref{2.10}), which can be shown by using the
conservation equation  (\ref{2.11}).
The line element can be written directly in terms of $p, K, trK, R$ and $N$:
\begin{equation}
ds^2= dt^2(-N^2+{N^2(trKR-KR)^2\over (pR)^2}) -2N{trK-K \over p}dtdR
+{4\over (pR)^2}dR^2 +R^2d\Omega^2.
\label{2.18}
\end{equation}
Equations (\ref{2.13} - \ref{2.18}) demonstrate that  all metric functions
can be expressed as certain functionals of matter-related terms. That is
simply a  manifestation of the well known fact  that spherically symmetric
gravitation does not carry
 degrees of freedom independent of matter.
The scalar field equation can be cast into one of two equivalent forms
 \begin{eqnarray}
D\underline D\phi = {-\theta '\over 2}D\phi -  {\theta '\over 2}\underline D
\phi -{\partial_rN\over N\sqrt{a}}\underline D\phi
-trK \underline D\phi +W',
\label{2.19}
\end{eqnarray}
 \begin{eqnarray}
\underline DD \phi = {-\theta \over 2}D\phi -  {\theta \over 2}\underline D
\phi -{\partial_rN\over N\sqrt{a}}  D\phi + trK   D\phi +W' .
  \label{2.20}
\end{eqnarray}
 Here the differential operators $D, \underline D $ are defined as
follows
 \begin{eqnarray}
D = { 1\over N}\partial_t +  { 1\over \sqrt{a}}\partial_r  ~~~~
\underline D = { -1\over N}\partial_t +  { 1\over \sqrt{a}}\partial_r,
\label{2.21}
\end{eqnarray}
  $\theta '=p+K -trK,~~\theta =p-K+trK $  are the optical scalars
  and $W'=\partial_{\phi }W(\phi )$. $W(\phi )$ is the scalar field
selfinteraction potential.

The nonzero components of the energy-momentum  tensor of the scalar field
are
 \begin{eqnarray}
&&T_0^0 ={-1\over 4}((D\phi )^2+ (\underline D\phi )^2+4W(\phi )   ),~~~
T^r_r ={1\over 4}((D\phi )^2+ (\underline D\phi )^2- 4W(\phi ) )
\nonumber\\ &&
{T^0_r\over \sqrt{a}}={-1\over 4N}((D\phi )^2- (\underline D\phi )^2 ),~~~~~
T_{\phi }^{ \phi }=T_{\theta }^{ \theta } ={-1\over 2}( D\phi \underline
D\phi  - 2W(\phi )).
\nonumber\\ &&
\label{2.22}
\end{eqnarray}
  \vskip 2cm

\centerline{3.   Definition of the external Cauchy problem.}

We will deal with an external Cauchy problem with a free inner boundary.
Define  $\Sigma_0^{out}$ as an open end  extending
  outside a  sphere of a coordinate radius $R_0$.

 Initial  data entirely determine the geometry of the initial
slice and, given the conservation equation of the energy momentum tensor,
the lapse function $N$; see equations (\ref{2.13} - \ref{2.20}).
That specifies also null cones attached at $\Sigma_0$.
 Take now  an outgoing  null cone $H$ originating at a  radius $R_0$.
The open end  $\Sigma_0^{out}$   shall give rise to a foliation defined by
open Cauchy ends $\Sigma_t^{out}$; notice, however, that the foliation
is nonunique in general.  That is because on all future ($t>0$) open ends
$\Sigma_t^{out}$ the extrinsic curvature $K$ (see (\ref{2.13})) depends on a
parametr $C$ that is arbitrary. One would have to impose a new condition
on the free inner boundary
(in addition to the fixing of the trace of extrinsic curvature $trK$ of
$\Sigma_t^{out}$) that should
guarantee that the lapse function $N$ is strictly positive
in $\Sigma_t^{out}$. This is in order to guarantee that the proper time
$\tau =\int Ndt$ runs forward in the outer region.  This demand is quite
restrictive, in general.
>From formula (\ref{2.16})  follows that the lapse can vanish on a centered
sphere $S$ if the mean curvature $p(S)=0$, but it might vanish also
elsewhere, if  the second factor  of (\ref{2.16}) equals 0. The second
posibility is more difficult to deal with, since the factor in question
depends on the acceleration of matter (see (\ref{2.11}) and (\ref{2.13})).
Fortunately, the arising difficulty can be avoided by imposing  the polar
gauge condition $trK=K$; that condition removes   the arbitrariness
that is present in other gauges so that there is no free data at the inner
boundary.
Indeed, in such a case equations (\ref{2.14} -\ref{2.18}) become
\begin{equation}
pR=2\sqrt{1-{2m\over R}+{2m(R)\over R}},
\label{3.1}
\end{equation}
\begin{equation}
m(R)= 4\pi \int_R^{\infty }dr r^2\rho ,
\label{3.2}
\end{equation}
\begin{equation}
N={pR\over 2} \beta (R),
\label{3.3}
\end{equation}
\begin{equation}
\beta (r) = e^{16\pi \int_r^{\infty }(-T_r^r-\rho )
{1\over p^2s}ds}.
\label{3.4}
\end{equation}
and
\begin{equation}
ds^2= -dt^2N^2  +{4\over (pR)^2}dR^2 +R^2d\Omega^2.
\label{3.5}
\end{equation}
The parameter $m$ appearing above is a free parameter that will be assumed
to satisfy the inequality $m\ge m(R_0)$; the equality might take place 
only if the energy  density $\rho $ vanishes at the inner boundary
(exact conditions are given in Theorems 1 and 2).
(\ref{3.1}) -  (\ref{3.4}) imply that $N$ is positive provided that $p >0$.
 The momentum constraint  (\ref{2.8}) can be used  in order to replace  $trK$
by ${-8\pi \over p}j_r/\sqrt{a}$  and the equation (\ref{2.9}) allows one to
eliminate the gradient $\partial_rN$. With all that equations (\ref{2.19})
and  (\ref{2.20})  lead to

\begin{eqnarray}
(\partial_0 +{NpR\over 2}\partial_R)V = {8\pi N\over p}V(j-T) -
{Np\over 2}(U+V/2) -{NV\over pR^2} +NW'(\phi ),
\label{3.6}
\end{eqnarray}
 \begin{eqnarray}
(\partial_0 -{NpR\over 2}\partial_R)U = {8\pi N\over p}U(j+T) +
{Np\over 2}(V+U/2) +{NU\over pR^2} -NW'(\phi );
\label{3.7}
\end{eqnarray}
above we set
\begin{eqnarray}
&&V=\underline D\phi \nonumber\\
&&U=D \phi \nonumber\\
&&j={j_r\over \sqrt{a}}\nonumber\\
&&T=T_r^r.\nonumber\\
\label{3.8}
\end{eqnarray}

We will  refer to (\ref{3.6}) and (\ref{3.7})
as to "reduced Einstein-scalar field equations".
The whole dynamics of the
selfgravitating scalar field is given by the scalar field
equations (\ref{3.6}) and  (\ref{3.7}).
The scalar field $\phi $ can be written as an integral quantity
$\phi (R,t)=\phi (R,0)+\int_0^td\tau N(U-V)/2$ so that functions $U$ and $V$
can be used as dynamical variables.
The initial data of the whole system of Einstein - scalar field equations
consist of $U$ and $V$ supplemented  by a value of $\phi $ at a single
point (say $\phi (t=0, \infty )=0$).

Equations (\ref{3.6}) and (\ref{3.7}) are strictly hyperbolic  in the sense
of Petrovsky provided that $NpR$ is
strictly  positive. One can show that $NpR$ can vanish only if $pR=0$. 
On the other hand,   $\partial_r\beta $  (and hence some of coefficients of
the reduced equations) can become singular at those points, where the mean
curvature $p$ vanishes. 
Thus in our case  strict  hyperbolicity is a prerequisite for the existence of
a local causal  evolution. 
That means, from the inspection of (\ref{3.1} and
(\ref{3.3}), that close to the sphere with $R=2m$ the contribution $m(R)$
has to be positive, that is guaranteed if $\rho \ge 0$; the latter condition
is satisfied if the selfinteraction $W(\phi )$ of the scalar field is
nonnegative. 
 Thus,  even the local version
of the Cauchy problem might require the positivity  of the potential
   term $W(\phi )\ge 0$.

As pointed above, the system of equations (\ref{3.6}) and (\ref{3.7}) is
strictly hyperbolic
in a region that does not contain minimal surfaces. From the inspection
of formulae for $p$ and $N$ it is obvious  that both quantities
are strongly positive outside
a sphere of  the Schwarzschild radius $R=2m$, so that outside that
region the information propagates   causally and time runs forward.
That would mean that the  sort of
the external Cauchy problem that is described in the beginning of this
Section is legitimate, at least for  initial open ends $\Sigma_0^{out}$
originating out of the sphere $R=2m$. We will show that to be true even in
a more general situation. We shall stress, however, that the  coordinates
that we use do not allow for the investigation of regions inside apparent
horizons because in the polar gauge  apparent horizons coincide with minimal
surfaces, where our system of coordinates breaks down.

 \vskip 2cm

\centerline{ 4. The  Cauchy solution.}

The reduced equations (\ref{3.6}, \ref{3.7}) are integro-differential
and hyperbolic; notice that they  are nonlocal.
I did not find any mathematical result  that can yield directly the existence of
a local in time solution for equations of that type. For that reason
we have to prove the existence of a local evolution starting
from first principles \cite{Choquet-Bruhat}.

There are several methods to prove the existence of a local Cauchy solution.
We will use an approach that bases on results of Petrovsky
\cite{Pietrovski} and on properties of Sobolev spaces.
The main result is formulated in Theorem 1. Its proof consists of
following main points.
Firstly, a  sequence of functions will be generated iteratively for data given on an
extended initial hypersurface.
Secondly, Lemma 3, proven
in the next Section, shows that  the sequence is uniformly bounded
for a short period of time outside a "rigid" cone. 
Then, in Lemma 4,  standard compactness
theorems of functional analysis  ensure  the  existence of a
 convergent  subsequence, that is  the sought local in time  solution of the
 reduced equations.  Finally, Lemma 7
 shows the existence of a global Cauchy evolution.

Let us recall that Sobolev spaces $H_k(V)$ can be defined as a completion
of $C^k$-functions in the norm $||f||_{H_k(V)}=
\int_VdV\Sigma_{i=0}^k(D^if)^2$ where
$D^if=\Sigma_{k_1,k_2...k_n: k_1+k_2+...k_n=i}\partial_{k_1}...
\partial_{k_n}f$ and $n$ is the dimension of a riemannian manifold $V$.
In the case of spherical symmetry the norm reads  $||f||=
\int drr^2\Sigma_{i=0}^k(\partial_r^if)^2$.
The main result is the following one.

{\bf Theorem 1.} Let the initial data of the reduced   equations (\ref{3.6} )
and (\ref{3.7}) be  $U, V\epsilon H_k$ and $\phi \epsilon H_{k+1}
\cap L_{n_0}$, k=2, 3,
... .
 Assume that $W(x)>0$ for $x\ne 0$ and $W(0)=0$ and  
$$  |\partial_x^lW(x)|\le A(l, n_l)|x|^{n_l}$$
for some $0\le n_l< \infty $ and a constant $A$ depending only on $n_l$ and
 $l=0, ..., k$. 
 Let  $\Sigma_0^{out}$ be an open end, $R_0=\inf [|{\bf x}|: {\bf x} 
 \epsilon \Sigma_0^{out}] > 2m +\eta $, $\eta >0$ and 
   $m \ge m(R_0)$ be an asymptotic mass of the configuration (with  strict
inequality $m > m(R_0)$ 
if  $\partial^i_Rf \ne 0$, $f=U, V$, 
at least for one value $i=0,...k-1$).
  
Then

i) there exists a local Cauchy evolution of $\Sigma_0^{out}$, i. e.
a foliation $\Sigma_t^{out}$ for some $\eta '$( $0\le t <\eta '$), with $U,
V\epsilon  H_{k}$ and  $\phi \epsilon H_{k+1}$.
 In addition, $U, V  \epsilon C^{k-1}_- (\bigcup_{0\le t<\eta '}\Sigma_t^{out})$;

ii) solutions are unique;

iii) solutions are global in $H_k$ and
 the null boundary $\delta H$ of
$\bigcup_{0\le t<\eta '}\Sigma_t^{out}$ escapes to spatial infinity,
i. e., the area of the inner boundary  $\Sigma_t^{out}$ goes to infinity as
 $t\rightarrow \infty $.

Actually, one can prove  a sligthly stronger version:

{\bf Theorem 2.} Let the initial data of the reduced   equations (\ref{3.6} )
 be  $U, V\epsilon H_k$ and $\phi \epsilon H_{k+1}\cap L_{n_0}$, 
 k=2, 3,....
Assume that  $W(x)>0$ for $x\ne 0$ and $W(0)=0$ and 
$$ |\partial_x^lW(x)|\le A(l, n_l)|x|^{n_l}$$
for some $0\le n_l< \infty $ and a constant $A$ depending only on $n_l$
and $l=0, ..., k$.
 Let  $\Sigma_0^{out}$ be an open end, $R_0=\inf [|{\bf x}|: {\bf x} 
 \epsilon \Sigma_0^{out}]  $ and 
   $m \ge m(R_0)$ be an asymptotic mass of the configuration (with  strict
inequality $m > m(R_0)$ if  $ \partial^i_Rf \ne 0$, $f=U, V$, 
at least for one value $i=0,...k-1$).

Assume that minimal surfaces are absent in $\Sigma_0^{out}$.
Then

i) there exists a local Cauchy evolution of $\Sigma_0^{out}$, i. e.
a foliation $\Sigma_t^{out}$ for some $\eta '$( $0\le t <\eta '$), with $U,
V\epsilon H_{k}$ and  $\phi \epsilon H_{k+1}$.
In addition, $U, V  \epsilon C^{k-1}_- (\bigcup_{0\le t<\eta '}\Sigma_t^{out})$;

ii)  solutions are unique;

iii) solutions are global in $H_k$ and the null boundary $\delta H$ of
$\bigcup_{0\le t<\eta '}\Sigma_t^{out}$ either 

a) escapes to spatial infinity,
i. e., the area of the inner boundary $\Sigma_t^{out}$ goes to infinity as
 $t\rightarrow \infty $,

or

b) stabilizes at  the  radius $R=2m_B$,
where $m_B$ is the Bondi mass of a black hole.

{\bf Comments.}  The condition that $m $ may be equal to $m(R_0)$
only if a sufficient number of derivatives of $U$ and $V$
vanishes, is always satisfied
if initial data on $\Sigma^{out }_0$ are obtained by restriction from 
initial data on a whole slice (including the origin) $\Sigma_0$. 
Conditions $U, V\epsilon H_k(\Sigma_0)$ mean (through Sobolev
embeddings theorems) that $k-1$ derivatives of $U$ and $V$ are continuous.
 If $m=m(R_0)$ then 
 $U(R)$ and $V(R)$ must identically vanish for $R\le R_0$, and  therefore 
 also their derivatives up to the order $k-1$. The symbol $C^k_-$ means
a class of functions which are almost $C^k$, that is they are of H\"older
 class $C^{k-1 +\kappa } $ for any $0<\kappa <1$.  Let us point out that
 the smoothness of $U$ and $V$ can be improved to $C^{k-1/2}$, 
 at least for initial data of compact support, following
 an argument of \cite{MEnew}.

In what follows we will prove Theorem 1 in four steps and five lemmae;
the proof of two of them is placed in  separate sections.

Theorem 2 can be obtained by improving one of the elements
in the reasoning.

{\bf Proof.} Step  I of Theorem 1.

Define a $H_k$ extension 
 $\tilde \Sigma_0^{out}=[{\bf x}\epsilon \Sigma_0:2m+\eta -\delta \le 
 | {\bf x}| <\infty ]$  of $ \Sigma_0^{out}$ by
  assuming following initial data in the interval $[2m+\eta -\delta, 
 2m+\eta  ]$  
$$\partial_r^{k}f(R)=0,~~~~~~
\partial_r^{k-1}f(R)=\partial_r^{k-1}f(R_0),~~~~~~~  
\partial_r^{k-i}f(R)=\partial_r^{k-i}f(R_0) -\int_R^{R_0} 
dr\partial_r^{k-i+1}f(r),$$
where $f=U, V$ and $i= 2,...k$.
The initial values of the scalar field in the extension can   
be obtained from integrating $U$ and $V$, $\phi (R)= \phi (R_0) 
-\int_R^{R_0}dr{1\over Npr}(U+V)$. 
One can check (see Appendix) that $H_k$ norms  of $U$ and $V$ over 
the extended manifold  $\tilde \Sigma_0^{out}$ differ from 
  $H_k$ norms  of $U$ and $V$ over $  \Sigma_0^{out}$  by a term bounded above by $\delta CH_k(\Sigma_0^{out})$, where $C$ is a constant. Similarly,
the mass function $m(R)\le  m(R_0)+\delta CH_1(\Sigma_0^{out})$ for
$R\epsilon [2m+\eta -\delta ,2m+\eta ]$. Thus for $\delta $ sufficiently small
the extended manifold can have the same asymptotic mass $m$ and lie 
outside the Schwarzschild radius $2m$.

Let  us define  following quantities
\begin{eqnarray}
&&p_0(R,t)=p(R,t=0)
\nonumber\\ &&
N_0(R,t) =N(R,t=0)
\nonumber\\ &&
U_0(R, t)=U(R, t=0),
\nonumber\\ &&
V_0(R, t)=V(R, t=0),
\nonumber\\ &&
\phi_0 (R, t)=\phi (R, t=0)
 \label{4.1}
\end{eqnarray}
as being identically equal to the initial data.  Let $\rho_0$, $T_0$
and $j_0$ be  the initial values of the energy density $\rho $, $T$ and
$j$, respectively.
Define also, for $n\ge 1$, functions
\begin{eqnarray}
&&\alpha_n(R,t)=\sqrt{1-{2m-8\pi \int_R^{\infty }drr^2\rho_n (r, t)\over R}}
\nonumber\\ &&
\beta_n(R, t) = e^{-4\pi \int_R^{\infty }drr{(T_n+\rho_n)\over  \alpha^2_n}}
\nonumber\\
&&\rho_n  ={1\over 4}((U_n )^2+ (V_n )^2+ 4 W(\phi_n)   ),~~~~~
T_n ={1\over 4}((U_n)^2+ (V_n )^2- 4 W(\phi_n)  )
\nonumber\\ &&
j_n={1\over 4}((V_n )^2- (U_n )^2 ), \nonumber\\
&&\phi_n (t, R)=\phi (R,0)+\int_0^td\tau \alpha_{n-1}(\tau , R)\beta_{n-1}
(\tau , R)\Bigl( U_{n-1}(\tau , R)-
V_{n-1}(\tau , R)\Bigr) /2.
\label{4.2}
\end{eqnarray}
All formulae above are expressed in terms of $U_n$ and $V_n$ which (for $n>0$)
are defined as solutions of following evolution equations
 \begin{eqnarray}
&&(\partial_0 +\alpha_{n-1}^2\beta_{n-1}\partial_R)V_n = \nonumber\\
&&4\pi \beta_{n-1}RV_n
(j_n-T_n) -  {\alpha_{n-1}^2\beta_{n-1}\over R}(U_n+V_n/2) -
{\beta_{n-1}V_n\over 2R}
+\alpha_{n-1}\beta_{n-1} W'(\phi_{n} ),
\label{4.3}
\end{eqnarray}
 \begin{eqnarray}
&&(\partial_0 -\alpha_{n-1}^2\beta_{n-1}\partial_R)U_n = \nonumber\\
&&4\pi \beta_{n-1}RU_n
(j_n+T_n) +   {\alpha_{n-1}^2\beta_{n-1}\over R}(V_n+U_n/2) +
{\beta_{n-1}U_n\over 2R}
-\alpha_{n-1}\beta_{n-1} W'(\phi_{n} ),
\label{4.4}
\end{eqnarray}
with initial conditions $U_n(R,0)=U_0(R)$ and $V_n(R,0)=V_0(R)$.

  Define
 $\Omega_t=[(t', {\bf x}):~ 0\le t'\le t, ~|{\bf x}| \ge 2m+\eta -\delta
 +2t']$
 and $\tilde \Sigma_t^{out}=[x\epsilon \Omega_t: x^0=t]$. Define
\begin{equation}
E_{kn}(\tilde \Sigma_t^{out})=\int_{\tilde \Sigma_t^{out})}
\Bigl( |\partial_r^kU_n|^2 + |\partial_r^kV_n|^2 \Bigr) .
\label{4.5}
\end{equation}
The essential part of   step I is following.

{\bf Lemma 3.}
  Let the initial data of the reduced   equations (\ref{3.6} )
and (\ref{3.7}) be of compact support and
$U_0, V_0 \epsilon H_k(\tilde \Sigma_0^{out})$, $k\ge 2$), for
k=2, 3,... Assume that
$$W(x) \ge 0,~~~~|\partial_x^lW(x)|\le A(l, n_l)|x|^{n_l}|$$
for some $0\le n_l< \infty $ and a constant $A$ depending only on $n_l$ and
$l$.
Then, for a small  enough time $t$:

i) for each $n>0$ there exists a solution $(U_n, V_n)$ of equations
(\ref{4.3}, \ref{4.4}) in $\Omega_t$;

ii)Sobolev norms  $H_k(\tilde \Sigma_t^{out})$   of $U_n$ and $V_n$ are
uniformly bounded for $t'<t$,
\begin{eqnarray}
E_{ln}(\Sigma_t^{out})\le {1\over \Bigl( ( E_0^l)^{-l-2}-
C_lt\Bigr)^{1/(l+2)}  },
\label{4.6}
\end{eqnarray}
where $l=0, 1,...,k$, $E_0^l =E_{ln}(\tilde \Sigma_0^{out})$ is the
initial value of the norm $E_{ln}$ and $C_l$'s are some constants;

iii) coefficients  $\alpha_n, \beta_n , \phi_n$ of (\ref{4.3}, \ref{4.4}) 
are at least
$C^k$ as functions of $(t, R)$ with  a  norm
$|f|_{C^k}=\sup \Sigma_{k_1+k_2=0}^k |\partial_0^{k_1}\partial_R^{k_2}f|$
(where $f=\alpha_n , \beta_n$) uniformly bounded by
(up to a constant
coefficient depending only on initial data)
\begin{equation}
\Bigl(  E_{kn}(\Sigma_t^{out})\Bigr)^{2k} \le
{1\over \Bigl( ( E_0^k)^{-k-2}-
C_lt\Bigr)^{2k/(k+2)}  }
 \label{4.7}
\end{equation}
and, in particular, $\alpha^2\beta_n \le 2$;

iv) $L_2(\Sigma_t^{out})$ norms of $\partial_0^l\partial_R^mf$ ($f=U_1,V_1$
and $l>0, l+m\le k$)  are uniformly bounded;

v) $|\partial_R^l\phi_n(t, R)| \le C_3+C_4t$, where $C_3$ and $C_4$ are constants
depending only on initial data and $0<l\le k$.

\vskip 0.8cm
We postpone the proof of Lemma 3 to the next Section. Assuming that it holds
true one shows the following

{\bf Lemma 4}.  Assume  conditions of Theorem 1 and, in addition,
that initial data are    of compact support
and   $k\ge 2$. There exists
a solution  $(U, V)\epsilon H_{k}(\Omega_t)$ and 
$(U, V)\epsilon C^{k-1}(\Omega_t)$
of the reduced equations (\ref{3.6}, \ref{3.7}) in $\Omega_t$.

Proof of Lemma 4. Notice that there exists a compact set $\Omega_t$
that contains supports of all functions $(U_n, V_n)$.
>From iii) and iv) of Lemma 3, $U_n, V_n$ constitute a
bounded subset of $H_k(\Omega_t)$. Therefore there exists a subsequence
$U_{n_k}, V_{n_k}$ that is weakly convergent in $H_k(\Omega_t)$ and
strongly convergent in $H_{k-1}(\Omega_t)$ to a limit $(U, V)$. 
  $(U, V) $ is H:older $C^{\mu }$ continuous, hence $(U, V)$ is a limit
also in the $L_{\infty }$ norm.  (U, V)
solves, by the construction of the iteration, the reduced equations
(\ref{3.6}, \ref{3.7}) in $\Omega_t$. The solution belongs to
 $ H_{k}(\Omega_t)$ and, since  our problem is essentially two-dimensional, it 
is $C^{k-1+\kappa }(\Omega_t)$, for any $0<\kappa <1$, by the 
 Sobolev embedding theorem. 
 That ends the proof of Lemma 4.

{\bf Lemma 5.} Under conditions of Theorem 1 and for $k\ge 2$, a solution
$(U, V)$ is unique in $\Omega_t$.

The proof goes in a completely standard way \cite{Pietrovski} and we omit it.
\vskip 0.8cm
Step II of  the proof of Theorem 1.

We can  remove the compactness condition of the support of initial data.

{\bf Lemma 6.} Under conditions of Theorem 1
a solution $(U, V)$ with (noncompact) initial data that  are of class $ H_k$, k= 2, 3,...,
exists in $\Omega_t$.

Sketch of the proof of Lemma 6.

We shall discuss the case with $k=2$ only.
Take a $C^{\infty }$ sequence of functions of compact support $U_{0n}, V_{0n}$
that approximates $(U_0, V_0)$ in $H_2(\tilde \Sigma^{out}_0)$.  
Each pair of the sequence $U_{0n}, V_{0n}$ gives rise to a solution 
$U_{tn}, V_{tn}$ of the reduced equations (\ref{3.6}, \ref{3.7}) 
in $\Omega_t$.  From the reduced equations one derives, after a lengthy but 
simple algebra,
\begin{eqnarray}
&&(\partial_t+2\partial_R)\Bigl( |U_{tn}|_{H_2(\tilde \Sigma_t^{out})}^2 +
|V_{tn}|_{H_2(\tilde \Sigma_t^{out})}^2 \Bigr) \le \nonumber\\
&&C\Bigl( |U_{tn}|_{H_2(\tilde \Sigma_t^{out})}^2 +
|V_{tn}|_{H_2(\tilde \Sigma_t^{out})}^2 \Bigr)^3;
\label{4.8}
\end{eqnarray}
 Thus the sequence of solutions is uniformly bounded in $H_2(\Omega_t)$,
for sufficiently small but nonzero   $t > 0$.
Therefore there exists a subsequence $(U_{tn_k}, V_{tn_k})$ that is
weakly convergent to $(U, V)$ in $H_2(\Omega_t)$ and strongly
convergent in $H_1(\Omega_t^M)$, where 
$\Omega_t^M=[(R, t)\epsilon \Omega_t:~ R<M]$.  The last fact and the H\"older 
continuity of $(U, V) $ and $(U_{tn_k}, V_{tn_k})$ imply pointwise  
convergence of the subsequence in $\Omega_t^M$. A careful analysis 
of the reduced equation shows that for $M>>1$ the reduced equations are
nearly linear. A perturbation analysis shall prove the existence
of a solution in $\Omega_t \setminus \Omega_t^M $ that matches to
a solution $(U, V)$. That  concludes the
 proof of Lemma 6.
 \vskip 0.8cm

Step III of the proof of Theorem 1.

The inner boundary $\delta \Omega_t$ of  $ \Omega_t$   consists of pairs
$(t, 2m+\eta  -\delta +2t)$. The outgoing null rays  in $ \Omega_t$ are determined
from the differential equation ${dR\over dt}= {(pR)^2 \over 4}\beta $.
>From (\ref{3.1} - \ref{3.4}) we conclude that ${dR\over dt}<1$ at
$\delta \Omega_t$. From the smoothness condition we infer that
${(pR)^2\over 4} \beta^{1/2}$ is at least a bounded function along a null
ray. Thus $\delta \Omega_t$   is a space-like
three-surface. Therefore   
there exists a null outgoing  geodesic joining each
point $(t, R)$ (for small enough $t$)  of $\delta \Omega_t$ with  a point $P$
on  the initial open end  $\tilde \Sigma_0^{out}$. Let us choose $t$ such that
the point $P$ lies on the boundary of $\Sigma_0^{out}$; that is possible since
by  definition $\Sigma_0^{out}\subset \tilde \Sigma_0^{out}$. 
Then there exists
a time development $  \Sigma_t^{out}  \subset \tilde  \Sigma_t^{out}$,
defined by the solution $(U, V)$ of the reduced Einstein-scalar field
equations, of the initial open end $ \Sigma_0^{out}$, i. e. 
$ \Sigma_0^{out}$ is future null geodesically complete. That accomplishes 
the local proof of Theorem 1.
\vskip 0.8cm

Step IV of the proof of Theorem 1.

{\bf Lemma 7.}  Define $g={V\over pR}$ and $h={U\over pR}$.
Under conditions of Theorem 1,

Proposition 7.1.  Banach norms $L_{2j}$ of $g$ and $h$ are equivalent to
 norms $L_{2j}$ of $U$ and $V$, respectively;

Proposition 7.2. Sobolev norms $H_l$ and integral norms
$L_{2j}$ of $g$ and $h$ are uniformly bounded
by $Ce^{Ct}$ for any integer $j$, $l\le k$  and for some constant $C$;

Proposition 7.3.  The scalar field $\phi $ is uniformly bounded by
a constant $C$ depending only on an asymptotic (ADM) mass of the 
configuration and  on the areal radius $R_0$ of 
the initial slice $\Sigma_0^{out}$,
\begin{equation}
|\phi (R)| \le C.
\label{4.9}
\end{equation}
The proof of Lemma 7 is highly technical and we postpone it to
Section 6.  Using the above Propositions  we infer that, under conditions
of Theorem 1, there exists a  uniform bound on $L_{2j}$ norms of $U$ and
$V$. We show   that $H_k$ norms of $U$ and $V$ remain  bounded 
during a finite evolution. 
Indeed,
\begin{eqnarray}
&&||\partial_rU||^2_{L_2(\Sigma_t)}=\int_{\Sigma_t}dV\Bigl( \partial_r(prh)
\Bigr)^2\le  \nonumber\\
&&2 \int_{\Sigma_t}dV\Bigl( (\partial_r(pr)h)^2+ (pr\partial_rh)^2\Bigr) \le 
\nonumber\\ 
&&C  \int_{\Sigma_t}dV\Bigl( (U^2+V^2)^2  h^2+ (\partial_rh)^2\Bigr),
\label{4.10}
\end{eqnarray}
for some $C$.
The first inequality is just a  trivial application of
$2ab\le a^2+b^2$ and in the second one
we use $0<pr\le 2$, expression (\ref{3.1}) (with $\rho ={1\over 4}(U^2+V^2)  
+W(\phi )$) and Proposition 7.3.  The last expression of (\ref{4.10})
is globally bounded by applying  the Schwartz inequality and then using 
Propositions 7.1 and 7.2. Thus the norm $||\partial_rU||^2_{L_2(\Sigma_t)}$
is globally bounded. 

Similarly one shows the global boundedness of 
$||\partial_rV||^2_{L_2(\Sigma_t)}$ and, iteratively, of higher Sobolev
norms of functions $U$ and $V$.

One can show that the area of the inner boundary of $\Sigma_t $
 goes to   infinity as $t\rightarrow \infty $. Indeed, the inner  boundary
$\bigcup_t\Sigma_t $ is null and its areal radius $R_t$ satisfies the
differential equation   ${dR_t\over dt}= {(pR_t)^2 \over 4}\beta $.
Initially we have $R_0>2m+\eta $, with $\eta >0$; thus from (\ref{3.1}),
(\ref{3.4})  and the equation we infer that $(pR)^2(t)>4(1-{2m\over 2m+\eta })$
and $\beta (R_t) > e^{{-4\pi \over 2m(1-{2m\over 2m+\eta })}\int_{R_t}dr
r^2(\rho +T)} \ge e^{{- 1\over 2(1-{2m\over 2m+\eta })}}$. 
Thus ${dR_t\over dt}$
is strictly positive nad $R_t$ goes to infinity. 
 That completes the global part of Theorem 1.

\vskip 2cm
\centerline{5. Proof of Lemma 3. }

  The  approximating equations  (\ref{4.3}) and (\ref{4.4})
are nonlinear in $U_n, V_n$, with polynomial nonlinearities in each order
$n$, so that their coefficients are of class $C^{\infty }$ as functions of
  unknown variables $U_n, V_n$. A long but simple calculation yields
following equations
 \begin{eqnarray}
&&(\partial_0 + 2\partial_R)
\int_R^{\infty }dr {r^2\over 2}\Bigl( (\partial^k_RU_n)^2
+(\partial^k_RV_n)^2\Bigr) =
\nonumber\\
&&{R^2\over 2}\Bigl( (\partial^k_RU_n)^2 (-2-\alpha_{n-1}^2\beta_{n-1})+
(\partial^k_RV_n)^2 (-2+\alpha_{n-1}^2\beta_{n-1})\Bigr) \Big|_R+ \nonumber\\
&&\int_R^{\infty }drr^2 \Big[ \Gamma_n-{1\over 2r^2}\partial_r
(\alpha_{n-1}^2\beta_{n-1})+k\partial_r(\alpha_{n-1}^2\beta_{n-1})\Bigr)
\Bigl(  (\partial^k_rU_n)^2- (\partial^k_rV_n)^2 \Bigr)   + \nonumber\\
&&8\pi r\beta_{n-1}j_n   (\partial^k_rV_n)^2+ \partial^k_r
(\alpha_{n-1}\beta_{n-1}V') ( \partial_r^kV_n-\partial_r^kU_n)\Bigr) \Bigr]  +
\nonumber\\
&&\int_R^{\infty }dr r^2  \Sigma_{i=1}^k
\Bigl[ \Bigl( \partial_r^i\Gamma_n \Bigl( _i^k\Bigr) +
(1-\delta_{ki})\partial_r^{i+1}
(\alpha_{n-1}^2\beta_{n-1})\Bigl( _{i+1}^k\Bigr) \Bigr) \Bigl(
\partial_r^kU_n  \partial_r^{k-i}U_n-\partial_r^kV_n\partial_r^{k-i}V_n\Bigr)
+ \nonumber\\
&&8\pi \partial_r^i(\beta_{n-1}rj_n)\Bigl( _i^k\Bigr)\partial_r^kV_n
\partial_r^{k-i}V_n
+\Bigl( ^k_i\Bigr) \partial_r^i({\alpha_{n-1}^2\beta_{n-1}\over r})\Bigl(
\partial_r^{k-i}V_n  \partial_r^{k}U_n-\partial_r^{k-i}U_n\partial_r^{k}V_n
\Bigr)  \Bigr],
 \label{5.1}
\end{eqnarray}
where
\begin{equation}
\Gamma_n=4\pi \beta_{n-1}R(j_n+T_n)+{\alpha_{n-1}^2\beta_{n-1}\over 2R}
+{ \beta_{n-1}\over 2R},~~~\Bigl( ^k_i\Bigr)={k!\over i!(k-i)!}.
 \label{5.1a}
\end{equation}
In particular one obtains 
\begin{eqnarray}
&&\partial_0\int_R^{\infty }r^2dr\rho_n =
{1\over 4} (V_n^2-U_n^2)\alpha^2_{n-1}\beta_{n-1}R^2+
\nonumber\\
&&{ 1 \over 2} \int_R^{\infty }dr r^2 \beta_{n-1} \Bigl[ W'(\phi_n)
\Bigl( V_n-V_{n-1}-U_n+U_{n-1}\Bigr) +16\pi r j_n\Bigl( W(\phi_n) -
 W(\phi_{n-1}) \Bigr) \Bigr] .
 \label{5.1aa}
\end{eqnarray}

 Let us recall an obvious inequality
\begin{eqnarray}
| f(R)|\le {1\over R^{1/2}}
\Bigl( \int_R^{\infty } (\partial_Rf)^2r^2dr\Bigr)^{1/2}
\label{5.1c}
\end{eqnarray}

that holds true for any function of compact support that  belongs to $H_1$.

One gets from (\ref{5.1}), applying the Schwartz inequality and pointwise
estimates (\ref{5.1c}),  the inequality
\begin{eqnarray}
&&{d\over dt}E_{kn}\le C \Bigl[ E_{kn}\Bigl[ \sup |\Gamma_n|+
\sup \rho_{n-1}  +
\sup |j_{n}|\Bigr] +E^{1/2}_{kn} ||\partial_r^{k}
(\alpha_{n-1}\beta_{n-1}V')||_{L_2}+
\nonumber\\
&&\sum_{i=1}^{k-1}E^{1/2}_{k-in}E^{1/2}_{kn}\Bigl( \sup |\partial_r^i
\Gamma_n|  + \sup |\partial_r^{i+1}(\alpha_{n-1}^2\beta_{n-1})|\Bigr) +
\nonumber\\
&&E^{1/2}_{kn}E^{1/2}_{1n}\Bigl( ||\partial_r^k\Gamma_n||_{L_2}+
||\partial_r^{k+1}(\alpha_{n-1}^2\beta_{n-1})||_{L_2} \Bigr) \Bigr] .
\label{5.1d}
\end{eqnarray}
Here and below constants may change from a line to line, but they do depend
only on initial data and $k$.
Using (\ref{5.1c}) and various H\"older estimates in order to
bound the $\sup -$ terms on the left hand side of (\ref{5.1d})
one derives the differential inequality
\begin{equation}
(\partial_t+2\partial_R) E_{kn}(\Sigma_t^{out})\le
C_2E_{kn}\Bigl( E_{1n}^2+E_{1,n-1}+E_{k-1n-1}^k(E_{k-1n}+E_{kn-1})\Bigr) .
\label{5.2a}
\end{equation}
The interesting fact is that for $k>1$ the inequality (\ref{5.2a}) is
linear in the term $ E_{kn}$.

  I will use the method of exact induction in order to prove Lemma 3.
For $n=1$ the coefficients of
(\ref{4.3})  and \ref{4.4}) are time-independent.  Sobolev
embeddings theorem (or, strictly saying, (\ref{5.1c})) implies that $U_0$ and
$V_0$ are of H\"older class
$C^{k-1/2}$. Therefore the coefficients of  (\ref{4.2})
are of class $C^{k+1/2}$ for $R\ge R_0$ as functions of $R$ and (being
time-independent) they are
 bounded and  continuous in the  class $C^k$, $k\ge 2$ as functions of $t, R$.
The eigenvalue $\alpha_0\beta_0$ of the hyperbolic operator is bounded 
from above by 1 and it is strictly nonzero so that in the case $n=1$
the approximating equation is hyperbolic.  
A result of Petrovsky \cite{Pietrovski} ensures the existence of
$U_1, V_1\epsilon C^{k-1}(\Omega_t )$, confirming part i) of lemma 3.

   (\ref{5.2a}) becomes now
\begin{equation}
(\partial_t+2\partial_R) E_{k1}(\Sigma_t^{out})\le
C_2E_{k1}\Bigl( E_{11}^2+E_{1,0}+E_{k-10}^k(E_{k-11}+E_{k0})\Bigr) .
\label{5.2}
\end{equation}
where $E_{k0} =E_{kn}(\tilde \Sigma_0^{out})$  is the initial value of the norm
$E_{kn}$ and $C_1, C_2$ are some constants depending only on initial data.
The $C^{k-1}$ smoothness of the solution $U_1, V_1$ guarantees that $E_{l1}$
is a continuous function of time $t$ for $l<k$.
Now, let $E_{l1}(t) > E_0$ for   $t'\ge t \ge 0$. Then (\ref{5.2})
implies, after elementary calculations,
\begin{equation}
E_{l1}(\Sigma_t^{out})\le {1\over \Bigl( E_{l0}^{-l-2}-C_{l}
t\Bigr)^{1/(l+2)}  },
\label{5.3}
\end{equation}
 where   we used the initial value condition $E_{l1}=E_{l0}$. The above
 trick does not work for $E_{k1}$ since the latter need not be continuous.
 In order
 to show a corresponding  estimation for $E_{k1}$ one has to insert
 estimations (\ref{5.3}) into (\ref{5.2}) and integrate the latter.
The linearity of (\ref{5.2}) in $E_{k1}$ allows one to get an estimation as
in (\ref{5.3}).
 That  proves the point ii)  of Lemma 3.

In order to show iii) let us notice that $\partial^k_R\alpha_n$ or
$\partial^k_R\beta_n$ contain derivatives of $U_n$ and $V_n$ of the order
$k-1$ at most; that is   functions $\alpha_1$ and $\beta_1$  are of class $C^k$ for
fixed $t$. That can be shown quite generally, in any order.  
A direct calculation shows that
\begin{eqnarray}
\partial_0\alpha_n^2 =
{8\pi \over R}\partial_0 \int_R^{\infty }dr r^2\rho_n 
\label{5.4}
\end{eqnarray}
(\ref{5.4}) and (\ref{5.1aa}) show that  the order of differentiability of
$\partial_0\alpha_n$  is
equal to the order of differentiability of $U_n$ and $V_n$, if
coefficients of the preceding order, $\beta_{n-1}, \alpha_{n-1}$ are
of class $C^k$.
Therefore $\partial_0\alpha_1$ is $C^{k-1/2}$    in $\Sigma_t^{out}$
and $C^{k-1}$ in $\Omega_t$
because $U_1, V_1\epsilon C^{k-1/2}$ and they are $C^{k-1}$ in
$\Omega_t$.  One shows that
$$\partial_0 \beta_{n}=-4\pi \beta_n 
\int_R^{\infty }dr r^2 \Bigl[ 
-{2\rho_n^2\over r\alpha^4_n}\partial_0\alpha_n^2+$$
$$2j_n\alpha^2_{n-1}\beta_{n-1}\partial_r{1\over r\alpha_n^2} +$$
$${  \beta_{n-1} \over r\alpha^2_{n}}     \Bigl[ W'(\phi_n)
\Bigl( V_n-U_n\Bigr) +16\pi r j_n\Bigl( W(\phi_n) -
 W(\phi_{n-1}) \Bigr) \Bigr]  \Bigr]+$$
$${-8\pi \beta_nj_nR\over \alpha^2_{n}} \alpha^2_{n-1}\beta_{n-1} ; $$
obviously $\partial_0 \beta_{1}$  is $C^{k-1/2}$ in  $\Sigma_t^{out}$. 
Thus both $\alpha_1$ and
$\beta_1$ are at least $C^k$ in  $\Omega_t$. That accomplishes the first part
of the  statement iii). $C^{k-1}$ - differentiability  of $U_0$ and $V_0$
implies that also $\phi_1(t, R)=\phi_0(R)+\int_0^tds\alpha_0
\beta_0{U_0-V_0\over 2}$ is (at least) $C^{k }$. The estimate v)
is trivially true for $\phi_1$. The second part
of iii), the estimate (\ref{4.7}) and that of $\alpha_1^2\beta_1$,
can be obtained from $C^k$ estimates of $\alpha_1$ and $\beta_1$,
  expressing them  in terms of initial data (through (\ref{5.3})).

Differentiation of the approximating equations  (\ref{4.3}, \ref{4.4})
with respect $t$ and $R$, integration over $\Sigma_t^{out}$ and
the use of ii) and iii)
allows one to express  those   norms of functions $U_1$ and $V_1$ that are
specified in point iv)   in terms
of initial data.   That finishes the proof of the first step of the
induction hypothesis.

Now, let it be true for some $n$. Using the induction hypothesis
and the same reasoning as above, one shows that the coefficients of the
$n+1$ equation (\ref{4.3}, \ref{4.4}) are at least $C^k$, so that there
exists a solution
$U_{n+1}$ and $V_{n+1}$, by a result of Pietrovski (\cite{Pietrovski}).
The second power of its Sobolev norm, $E_{l,n+1}$ (where $l<k$), is either
less than $E_{l,n} $ for some $t$ close to $t=0$ or greater than
$E_{l,n} $.  In the former case it satisfies the bound
of (\ref{5.3}) by the induction hypothesis while in the latter case   we have,
from  (\ref{5.2a})
\begin{equation}
(\partial_t+2\partial_R) E_{l,n+1}(\tilde \Sigma_t^{out})\le
\Bigl( E_{l,n+1}(\tilde \Sigma_t^{out})\Bigr)^{l+1},
\label{5.5}
\end{equation}
which again yields
\begin{equation}
E_{l,n+1}(\tilde \Sigma_t^{out})\le {1\over \Bigl( E_{l0}^{-l-2}-
C_lt\Bigr)^{1/(l+2)}  }.
\label{5.6}
\end{equation}
The estimation of $E_{kn+1}$ can now be obtained in the same token as that
of $E_{k1}$.
Hence   i) and ii) are proven. The remaining steps, iii) and iv)  are
shown  identically as  in the case $n=1$.
We will check the validity of v).
Notice that
$$\phi_{n+1} =\phi_0(R)+\int_0^tds\alpha_n
 \beta_n{U_n-V_n\over 2}; $$
the right hand side of that equation is bounded from above by
$\sup |\phi_0(R)| +C\int_0^tdsE_{1n}^{1/2}$, using (\ref{5.1c}).
The induction hypothesis ii) and direct integration immediately yield v).
That ends the proof of Lemma 3. Let us point out that the uniform bounds 
proven above show the existence of all approximating functions in
a strictly positive time $t$; from (\ref{5.6}) follows that $t$ can be bounded
from below by a number that depends only on initial data and 
does not depend on $n$.

\vskip 2cm
\centerline{6. Proof of Lemma 7. }

Let $R_t=\inf_{\Sigma_t^{out}}(r)$; we have
\begin{equation}
R_t=R_0+\int_0^tds
(\partial_t+ {NpR\over 2}\partial_R)R=R_0+  \int_0^tds{NpR\over 2}\ge
R_0> 2m
\label{6.1}
\end{equation}
(these  inequalities follow from the assumption made in Theorem  1
that $\Sigma_0^{out}$  is placed outside a centered sphere of the
Schwarzschild radius $2m$).  Using (\ref{2.14}) and \ref{6.1}), we conclude
that
\begin{equation}
Rp(t,R)\ge
 2\sqrt{1-{2m\over R_0}}>0
\label{6.2}
\end{equation}
on all future slices  $\Sigma_t^{out}$.
Since (\ref{2.14}) implies also $Rp \le 2$, we  infer that there exist 
two nonzero and finite numbers $\alpha , \beta $ such that
$$\alpha g\le {V\over pR}\le \beta g,~~~\alpha h\le {U\over pR}\le \beta h;$$
that fact trivially implies the equivalence of $L_p$ norms of pairs
$(g,~V)$ and $(h,~U)$, for any p, and hence also for even values of $p$.
That proves Proposition 7.1.

One can show that
\begin{equation}
\partial_0(pR) = 8\pi NR j
\label{6.2a}
\end{equation}
The reduced equations (\ref{3.6}) and  (\ref{3.7}) lead now 
to following system
of equation for functions $g={V\over pR}$ and $h={U\over pR}$
\begin{eqnarray}
(\partial_0 +{NpR\over 2}\partial_R)g = 16 \pi {Ng\over p}W(\phi )
-{2N\over pR^2}g   -{Nph\over  2}  +{N\over pR}W'(\phi ),
\label{6.3}
\end{eqnarray}
\begin{eqnarray}
(\partial_0 -{NpR\over 2}\partial_R)h = -16 \pi {Nh\over p}W(\phi )
+{2N\over pR^2}h   +{Npg\over  2}  -{N\over pR}W'(\phi ).
\label{6.4}
\end{eqnarray}
The interesting fact is that these equations are almost linear in unknowns;
although   coefficients are some functionals of $g$ and $h$, they are 
pointwise bounded by constants that are independent of both functions.

Embeddings theorems imply that initial data $(U, V) $ that are of Sobolev 
class $H_k$ ($k>1$) must belong
to $L_p(\Sigma_t^{out})$ for any $p >2$; that implies, taking into account  
the definition of $g, h$ that initial data of the the latter are  
also of class $L_p$.
Let $n$ be an even number. Multiply (\ref{6.3}) and (\ref{6.4})  by  
$g^{n-1}$ and $h^{n-1}$, respectively and integrate over $\Sigma_t^{out}$.
Integrating by parts and using  (\ref{6.9}) (see below) one arrives at
\begin{eqnarray}
&&(\partial_0 +{NpR\over 2}\partial_R)\int_R^{\infty }dr r^2 {1\over n}
(h^n+g^n)= -{NpR^3\over n}h^n+\nonumber\\
&&\int_R^{\infty }dr r^2  \Bigl[ {2N\over pr}
 (g^n-h^n)  \Bigl(  {(pr)^2\over 4nr} +{1-n\over nr}+{8\pi (n-1)\over n}
rW(\phi )\Bigr) \nonumber\\
 &&gh{Np\over 2}(h^{n-2}-g^{n-2})
+{NW'(\phi )\over pr}(g^{n-1}-h^{n-1})\Bigr] .
\label{6.5}
\end{eqnarray}
The assumptions of Lemma 7 
$W(x)>0$ for $x\ne 0$ and $W(0)=0$ imply that for initial data of compact  
support the scalar field $\phi $ vanishes at spatial infinity; 
then one can use (\ref{5.1c}) and replace $\partial_r\phi $ by 
${ 1\over pR}(U+V)$  (see (\ref{3.8}), which leads finally
to the estimation 
\begin{equation}
\phi (R, t)\le C(||U||_{L_2(\Sigma_t^{out})}+ ||V||_{L_2(\Sigma_t^{out})})\le 
  C (||U||_{L_2(\Sigma_0^{out})}+ ||V||_{L_2(\Sigma_0^{out})};
\label{6.6}
\end{equation}
the last inequality is valid only if the selfinteraction term $W(\phi )$
is nonnegative and  it follows essentially from the conservation of 
asymptotic mass $m$.       The constant $C$ can be found explicitly  
and  the right hand side of (\ref{6.6}) can be found to be bounded by 
${2\over \sqrt{1- {2m\over R_0}}  }  \sqrt{{m\over \pi R_0}}$.  
With this we prove Proposition 7.3.

With the inequality (\ref{6.6}) and having hitherto known $L_{\infty }$
estimate  $pR\le 2$, we get from (\ref{6.5})
\begin{equation}
(\partial_0 +{NpR\over 2}\partial_R)\int_R^{\infty }dr r^2 {1\over n}
(h^n+g^n) \le C \int_R^{\infty }dr r^2 {1\over n}(h^n+g^n),
\label{6.7}
\end{equation}
where $C$ is a constant depending only on $n$ and initial data.
Therefore $L_n$ norms of $g$ and $h$ are bounded in a global time 
evolution.
That proves the $L_p$ part of Proposition 7.2.

Differentiating   equations (\ref{6.3}) and (\ref{6.4}) with respect
$r$, multiplying them   by $\partial_rg$ and $\partial_rh$, respectively, and integrating over
$\Sigma_t^{out}$ one arrives at
\begin{eqnarray}
&&{d\over dt}|_{out}\int_R^{\infty }dr r^2 {1\over 2}\Bigl( (\partial_rg)^2
+(\partial_rh)^2 \Bigr)=\nonumber\\
&&-{NpR^3\over 2}(\partial_rh)^2 +\int_R^{\infty }dr r^2\Bigl[  
 \Bigl( (\partial_rg)^2-(\partial_rh)^2\Bigr)
\Bigl( {-1\over 4}\partial_r(Npr)+16\pi {NW\over p} -{2N\over pr^2}
+{Npr^2\over 2}\Bigr)+ 
\nonumber\\
&&\partial_r\Bigl( {N\over pr}W'(\phi)\Bigl) (\partial_rg-\partial_rh)
+(g\partial_rg-h\partial_rh)\partial_r(16\pi {NW\over p}-{2N\over pr^2})
+{1\over 2}(g\partial_rh-h\partial_rg)\partial_r(Np)\Bigr]
\label{6.8}
\end{eqnarray}
Notice that  coefficients ${ 1\over 4}\partial_r(Npr)$, 
${16\pi NW\over p}+ {N\over pr^2}$ in front of $(\partial_rh)^2$ and 
$(\partial_rg)^2$ are pointwise bounded on $\Sigma_t^{out}$. 
That follows from obvious estimates $0< N, pR/2\le 1$,  $R>2m$, from
the equation
\begin{eqnarray}
&&{-r^3\over 4}\partial_r(Npr^{-1})= 
{-N\over pr}+{3Npr\over 4} -4\pi Np^{-1}r(T-\rho )= \nonumber\\
&&{-N\over pr}+{3Npr\over 4} +8\pi Nr W(\phi )/p
\label{6.9}
\end{eqnarray}
and from Proposition 7.3.
Using the Schwartz inequality, Proposition  7.3 and the already proven 
part of Proposition 7.2 one obtains from 
(\ref{6.8}) the inequality
\begin{eqnarray}
{d\over dt}|_{out}\int_R^{\infty }dr r^2 {1\over 2}\Bigl( (\partial_rg)^2
+(\partial_rh)^2 \Bigr)\le   C
 \int_R^{\infty }dr r^2 {1\over 2}\Bigl( (\partial_rg)^2
+(\partial_rh)^2 \Bigr) 
\label{6.10}
\end{eqnarray}
with a constant $C$ depending  smoothly on various norms $L_{2j}$ and on the 
estimation of the scalar field. From (\ref{6.10}) we infer that the  integral 
$ \int_R^{\infty }dr r^2 {1\over 2}\Bigl( (\partial_rg)^2
+(\partial_rh)^2 \Bigr) $  remains bounded during a finite evolution. 
Therefore $H_1$ norms of $U$ and $V$ remain bounded.

Now, let there exists a global estimate for the Sobolev norm $H_{k-1}$
of $g$ and $h$.
In one dimension (to which our problem essentially reduces) 
the boundedness of $H_k$ implies the existence
of $L_p$ estimates of $\partial_r^lf$, $f=V, U$, for $l<k$ and for
any integer $p>2$. Differentiating reduced equations (\ref{6.3}) and
(\ref{6.4}) $k$ times with respect $r$, integrating the resulting equations,
using various H\"older inequalities and the induction hypothesis, one
obtains an inequality analogous to (\ref{6.10}) for the $H_k$ norm.
That leads to the conclusion that $H_k$  norm is also globally
bounded. That ends the proof of Proposition 7.2.
   
   \vskip 2cm
\centerline{ 7.  Proof of Theorem 2.}

Now I will discuss shortly the proof of Theorem 2.
It  proceeds in a way analogous to the proof of Theorem 1.
In particular, as in the former case, one defines the approximation
procedure of Step 1 of Theorem 1.  The only difference between the two
cases is due to the fact that previously we used an open  initial end
$ \Sigma_0^{out}$ with  $|{\bf x}|> 2m$ while now  it penetrates a region
inside the Schwarzschild radius  $|{\bf x}|= 2m$. The difference is
important, since the former
choice guarantees strict hyperbolicity of all approximating equations (\ref{4.3},
\ref{4.4}) - all coeeficients $\alpha_n^2\beta_n$ are uniformly bounded
away from  0 - while now we have to show that. 
 
 Let $R_0<2m $ be the areal radius of the inner boundary of $\Sigma_0^{out}$.
 Define  $ R_1 =R_0-\tau >0$   and
 $\tilde \Sigma_0^{out}=[{\bf x}\epsilon \Sigma_0: R_1  \le | {\bf x}|
 <\infty ]$ (an open end containing $ \Sigma_0^{out}$). Let
 $S_r^{out}=[{\bf x}\epsilon \Sigma_0:  R_0\le r  \le | {\bf x}|
 <\infty ]$ be an open end contained in  $ \Sigma_0^{out}$.
 Define a four-manifold
 $\Omega_{t}=[(t', {\bf x}):~ 0\le t'\le t, ~|{\bf x}| \ge R_1+2t']$
 foliated by hypersurfaces  $\tilde \Sigma_t^{out}=[x\epsilon
 \Omega_{t}: x^0=t]$ and another four-manifold $\Omega_{S_rt}=[(t',
 {\bf x}):~0\le t'\le t, ~|{\bf x}| \ge  r-2t']$ with leaves $S_{rt}^{out}=
[x\epsilon \Omega_{rt}: x^0=t]$. Obviously, $S_{r0}^{out}=S_r^{out}$.

As in Section 4 we define energy norms of solutions of the approximating
equations (\ref{4.3}), \ref{4.4}) for each  family of foliations,
\begin{equation}
E_{kn}^{\Sigma }(\tilde \Sigma_{t}^{out})= \Bigl( |\partial_r^kU_n
|_{L_2(\tilde
\Sigma_{t}^{out})}^2 + |\partial_r^kV_n|_{L_2(\tilde \Sigma_{t}^{out})}^2
\Bigr) ,
\label{7.1a}
\end{equation}
\begin{equation}
E_{kn}^{S_r }(\tilde S_{rt}^{out})= \Bigl( |\partial_r^kU_n
|_{L_2(\tilde
S_{rt}^{out})}^2 + |\partial_r^kV_n|_{L_2(\tilde S_{rt}^{out})}^2 \Bigr) .
\label{7.1b}
\end{equation}

    All iterative solutions satisfy the initial data,
$U_n(t, R)=U_0(R)$,  $V_n(t, R)=V_0(R)$;   norms of initial data are
denoted simply as $E_{k0}$ and $E^{S_r}_{k0}$ respectively,
for each of the two families of foliations.

The main technical result of Section 4, Lemma 3, has to be replaced by

{\bf Lemma 3'.}
  Let the initial data of the approximating   equations (\ref{3.6} )
and (\ref{3.7}) be of compact support and  let there exists a nonzero $\tau $
such that
\begin{equation}
 \inf_{R_0\le r\le \infty }\Bigl[ E_{0}^{S_r}-(2m-r+2\tau )\Bigr] >C_6\tau ,
\label{7.0}
\end{equation}
where $C_6$ is a constant defined below that depends  on initial data.
  Let
$U_0, V_0 \epsilon H_k(\tilde \Sigma_0^{out})$, $k\ge 2$), for
k=2, 3,... Assume that  $W(x) > 0$ if $x\ne 0$ and $V(0)=0$ and
$$(|\partial_x^lW(x)|\le A(l, n_l)|x|^{n_l}|$$
for some $0\le n_l< \infty $ and a constant $A$ depending only on $n_l$ and
$l$.
Then  for $t<\tau $

O)  equations (\ref{4.3}) and (\ref{4.4}) are strictly 
hyperbolic in $\Omega_{t}$;

i) for each $n>0$ there exists a solution $(U_n, V_n)$ of equations
(\ref{4.3}, \ref{4.4}) in $\Omega_{1t}$;

ii)Sobolev norms  $H_k(\tilde \Sigma_t^{out})$   of $U_n$ and $V_n$ are
uniformly bounded,
\begin{eqnarray}
E_{ln}^{\Sigma }(\tilde \Sigma_t^{out})\le {1\over \Bigl( ( E_{l0})^{-l-2}-
C_lt\Bigr)^{1/(l+2)}  },
\label{7.2}
\end{eqnarray}
where $l=0, 1,...,k$ and $C_l$'s are some constants;

iii) coefficients  $\alpha_n, \beta_n$ of (\ref{4.3}, \ref{4.4}) are at least
$C^k$ as functions of $(t, R)$ with  a  norm
$|f|_{C^k}=\sup \Sigma_{k_1+k_2=0}^k |\partial_0^{k_1}\partial_R^{k_2}f|$
(where $f=\alpha_n , \beta_n$) uniformly bounded by
(up to a constant coefficient depending only on initial data)
\begin{equation}
\Bigl(  E_{kn}^{\Sigma }(\tilde \Sigma_t^{out})\Bigr)^{2k} \le
{1\over \Bigl( ( E_{k0})^{-k-2}-
C_lt\Bigr)^{2k/(k+2)}  },
 \label{7.3}
\end{equation}

iv) $L_2(\tilde \Sigma_{t}^{out})$ norms of $\partial_0^l\partial_R^mf$
($f=U_n,V_n$ and $l>0, l+m\le k$)  are uniformly bounded;

v) $|\partial_R^l\phi_n(t, R)| \le C_3+C_4t$, where $C_3$ and $C_4$ are
constants depending only on $l$ and initial data and $0<l\le k$.

{\bf Proof.} The only new element in comparison with Lemma 3 is the
strict hyperbolicity assertion, point O above. We have to show that the
product $\alpha_n^2\beta_n$ is strongly positive for sufficiently small
values of $\tau $. That will be achieved if we show that $\alpha_n$ itself
is positive, that is (from (\ref{4.2}))
\begin{equation}
R+2t -2m + 8\pi \int_{R+t}^{\infty }dr r^2 \rho_n(r,t) >0
\label{7.4}
\end{equation}
for $R \ge R_1$.
That is obviously true for the $n=1$ equation, since all  functions
with a suffix "0"  (including $\alpha_0$)   are time-independent and
initially $\alpha $ is positive.
Now let the $n-th$ step of the induction hypothesis be true. We shall show
that  the $(n+1)th $ equations are strictly hyperbolic.
Manipulating with equations (\ref{4.3}) and (\ref{4.4})  one  finds the
 equation
 \begin{eqnarray}
&&(\partial_0 - 2\partial_R)2 \int_{R }^{\infty }dr r^2 \rho_n(r,t) =
(\partial_0 - 2 \partial_R)
\int_R^{\infty }dr {r^2\over 2}\Bigl( ( U_n)^2
+( V_n)^2 +2W(\phi_n) \Bigr) =
\nonumber\\
&&{R^2\over 2}\Bigl( ( U_n)^2 (2-\alpha_{n-1}^2\beta_{n-1})+
( V_n)^2 (2+\alpha_{n-1}^2\beta_{n-1})\Bigr) \Big|_R+ \nonumber\\
&&\int_R^{\infty }drr^2 \Big[ 8\pi \beta_{n-1} rj_n  (\rho_n-\rho_{n-1}
+T_{n-1}-T_n )+V'
\alpha_{n-1} \beta_{n-1}(U_{n-1}  -U_n + V_n-V_{n-1})  \Bigr]
\nonumber\\
\label{7.5}
\end{eqnarray}
that describes the evolution of the  $n-th$ order  external  energy (modulo a
coefficient $4\pi $) $\int_{R }^{\infty }dr r^2 \rho_n(r,t) $ along the
foliation   formed by leaves $S_{Rt}^{out}$.
>From that we obtain (dropping out the positive boundary term and
using pointwise estimates   in order to bound
the last  integral of (\ref{7.5}))
 \begin{eqnarray}
&&(\partial_0 - 2\partial_R) 2 \int_{R }^{\infty }dr r^2 \rho_n(r,t) \ge
 \nonumber\\
 &&\int_R^{\infty }drr^2 \Big[ 8\pi \beta_{n-1} rj_n  (\rho_n-\rho_{n-1}
+T_{n-1}-T_n )+V'
\alpha_{n-1} \beta_{n-1}(U_{n-1}  -U_n + V_n-V_{n-1})  \Bigr] \ge \nonumber\\
&&-C\Big( \sup (|j_n| + |V_n|+|U_n|+|V_{n-1}|+|U_{n-1}| )\Bigr)
\int_{R }^{\infty }dr r^2 \rho_n(r,t),
\label{7.6}
\end{eqnarray}
where $C$ is a constant.
Using now (\ref{5.1c})  and (\ref{4.5}) we observe  that $\sup
(|j_n| + |V_n|+|U_n|+|V_{n-1}|+
|U_{n-1}|$ can be bounded from above by $\tilde C_1 E_{1n}^{\Sigma }(\tilde
\Sigma_{1t}^{out})+\tilde C_2 \Bigl( E_{1n}^{\Sigma }(\tilde \Sigma_t^{out})
\Bigr)^{1/2}$, where $\tilde C_1$ and $\tilde C_2$ are constants. Using now
the induction
assumption (\ref{4.6}) and using the obvious inequality $\sqrt{x}+x\le 1+2x$,
we arrive at
 \begin{eqnarray}
&&(\partial_0 - 2\partial_R) 2 \int_{R}^{\infty }dr r^2 \rho_n(r,t) \ge
 \nonumber\\
  && -  C_4\Bigl( 1+{1\over \Bigl( E_{10}^{-3}-
C_1t\Bigr)^{1/3}  } \Bigr) 
  \int_{R}^{\infty }dr r^2 \rho_n(r,t),
\label{7.7}
\end{eqnarray}
which yields
\begin{eqnarray}
&&\int_{R-2t}^{\infty }dr r^2 \rho_n(r,t) \ge  e^{-C_5t}
\int_{R}^{\infty }dr r^2 \rho (r,t=0) \ge  \nonumber\\
\int_{R}^{\infty }dr r^2 \rho (r,t=0)-C_6t.
\label{7.8}
\end{eqnarray}
All constants above depend only on initial values and the degree of
nonlinearity of the potential term $W(\phi_n)$. The last constant enters
the condition (\ref{7.0}) stated in Lemma 3'.
 Using (\ref{7.8}) we can bound
the left hand side of (\ref{7.4}) as follows
\begin{eqnarray}
&&R+2t  -2m + 8\pi \int_{R+2t }^{\infty }dr r^2 \rho_n(r,t)\ge
 \nonumber\\
&&R  -2m +
8\pi \int_{R+4t}^{\infty }dr r^2 \rho (r,t=0)+ t (2-C_6).
\label{7.9}
\end{eqnarray}
Take now $  \tau $ of (\ref{7.0}) and notice that the condition
(\ref{7.0}) is satisfied by all values $0\le t\le \tau $.
Notice that $E_0^{S_{R+4t}} = 8\pi \int_{R+4t}^{\infty }dr r^2 \rho (r,t=0)$;
inserting  the condition $E_{0}^{S_{R+4t }}-(2m-(R+4t)+2\tau ) >C_6t  >0$
of (\ref{7.0}) into  (\ref{7.9}) we
 obtain  $(\tau -  t )(2+C_6)$, that is the last expression
of (\ref{7.9}) is positive for $t<\tau $. Thus we arrive at the desired  
result
$\alpha_n^2 (R, t)=R+2t  -2m + 8\pi \int_{R+2t }^{\infty }dr r^2 \rho_n(r,t) >
0$. The approximating  equations ($\ref{4.3}$) and (\ref{4.4}) 
are strictly hyperbolic in $\Omega_{t}$.

The rest of the proof of Lemma 3' is   similar to  that of Lemma 3.
 One can prove also results corresponding to those of lemmae 4-7, thus
accomplishing the proof of a local part of  Theorem 2 and the
first half of its global part.
I omit  details.  The proof of the remaining  global statements of the point
 iii) of Theorem 2 is given elsewhere \cite{MEnew}. Below I will sketch
the main idea. 

 Assume that we have a complete initial Cauchy hypersurface with an
apparent horizon, that is (using the polar gauge) with a minimal 2-surface
on the initial slice.  As a side remark let us point out 
that the area of an outermost  
apparent horizon cannot decrease (see, e. g., a proof outlined in 
\cite{MOM1994a}); in fact it has 
to increase whenever matter crosses through the horizon, which
moves acausally outwards. Asymptotically the spherically 
symmetric apparent horizont  becomes null and its   areal radius
 stabilizes at  $2m_B\le 2m$, where
$m_B$ is the Bondi mass.  Thus all  null geodesics starting outward
from $R\ge 2m_B$ at $t=0$ are complete.  The biggest set $H$ that is still 
future null complete can be constructed as follows.
Take a part $\Sigma_r$  of the initial hypersurface  that
does not include minimal surfaces. Then  data on  $\Sigma_r$ give rise to 
a local evolution, according to the local part of Theorem 2.   The global 
evolution prolongs  until the free  inner boundary  "freezes" 
 near a  minimal surface; the lapse collapses to  values close to 0
 and the area of the inner boundary is practically  constant. 
In  such a case one can take a slighly smaller initial open end  
$\Sigma_{r'}\subset
\Sigma_r$; that evolves to a spacetime that  freezes at a later time than
the previous one. Continuing that procedure ad infinitum  one finds finally
an open end such that a corresponding spacetime $H$ exists globally and  the 
area $dH$ of its null inner boundary  stabilizes at a finite 
value $4\pi R_B^2$ for $t\rightarrow \infty $.   
That limiting inner boundary $dH$ is an event horizon 
and half of $R_B$ is the Bondi mass.  It follows from the construction that 
all null geodesics originating in $H$ and directed outward are  complete.

Acknowledgements.  This research is supported by the KBN grant
2PO3B 090 08. 
I am  grateful to   Konstanty Holly and 
Andrzej Lasota from the Institute of  Mathematics of Jagellonian 
University and Atsushi Yoshikawa from the  Department of Applied 
Sciences of Kyushu University for their comments and useful suggestions.
I thank  Alan  Rendall from the Max Planck Institute in Potsdam
for many discussions during my visit in Potsdam and for pointing 
certain inconsistency in one of earlier versions of this paper.

\vskip 2cm
\centerline{Appendix}

{\bf Lemma.}

Assume notation of the main text. Let $U, V\epsilon H_2( \Sigma_0^{out})$.
Define a $H_k$ extension 
 $\tilde \Sigma_0^{out}=[{\bf x}\epsilon \Sigma_0:2m+\eta -\delta \le 
 | {\bf x}| <\infty ]$  of $ \Sigma_0^{out}$ by
  assuming following initial data in the interval $[2m+\eta -\delta, 
 2m+\eta  ]$  
$$\partial_r^{k}f(R)=0,~~~~~~
\partial_r^{k-1}f(R)=\partial_r^{k-1}f(R_0),~~~~~~~  
\partial_r^{k-i}f(R)=\partial_r^{k-i}f(R_0) -\int_R^{R_0} 
dr\partial_r^{k-i+1}f(r),$$
where $f=U, V$ and $i= 2,...k$.
Then 
$$||f||_{H_k(\tilde \Sigma_0^{out})}\le (1+C\delta ) ||f||_{H_k( 
\Sigma_0^{out})},$$
for sufficiently small $\delta $ and some constant $C$.

{Proof.} The two norms in question differ by a sum of terms of
the form $\int_{R_0-\delta }^{R_0}(\partial_r^{k-1}f(R))^2R^2dR$;
that sum is a polynomial in $\delta $ that can be bounded from above
by $C\delta \sum_{i/1}^{k-1}(\partial_r^{i}f(R_0))^2$. By  one of 
Sobolev inequalities (see the proof of Proposition 7.3), 
$|\partial_r^if(R_0)|\le C'||\partial_r^{i+1}f||_{  \Sigma_0^{out}}$;
combining    all information one arrives at the desired inequality.

The initial values of the scalar field in the extension can   
be obtained from integrating $U$ and $V$, $\phi (R)= \phi (R_0) 
-\int_R^{R_0}dr{1\over pr}(U+V)$.  That would lead to an estimation
of the contribution of the potential term to the mass $m$ that is due to the 
enlargement of the initial open end. Together with the preceding Lemma, 
that would give $m(R_0-\delta )\le m$ for sufficiently small $\delta $.


\begin{references}
\bibitem{Penrose} R. Penrose,  {\it Riv. N. Cimento}, {\bf 1}, 252(1969);
p. 631 - 668 in Seminar on Differential Geometry, Princeton: Princeton
University Press 1982;
\bibitem{Tipleratal} F. J. Tipler, C. J. S. Clarke and G. F. R. Ellis, p. 97
 in  General Relativity and Gravitation, vol. 2, ed. A. Held
New York 1980;
\bibitem{Wald} R. Wald,  General Relativity, Chicago: University of Chicago
Press 1984;
\bibitem{Leray} J. Leray, Hyperbolic Differential Equations, Princeton 1953;
\bibitem{Pietrovski} I. G.  Petrovsky, Lectures on Partial Differential
Equations, Warsaw 1955, PWN ( that is in Polish ; there are also
numerous Russian and English editions);
\bibitem{Penrose 1965} R. Penrose,  {\it Phys. Rev Lett.}, {\bf 14}, 57(1965);
 S. W. Hawking and G. F. R. Ellis, The Large Scale Structure
of Space-Time, Cambridge: Cambridge University Press;
\bibitem{Krolak} A. Kr\'olak, {\it  Class. Quantum Grav.} {\bf 3},
 267(1986); {\it J. Math. Phys.} {\bf 34}, 701(1992);
\bibitem{Moncrief} D. Eardley and V. Moncrief, {\it Gen. Rel. Gravitat.}
{\bf 13}, 887(1981);
\bibitem{Chrusciel}; P. T. Chru\'sciel, J. Isenberg and V. Moncrief,
{\it Class. Quantum Grav.} {\bf 7}, 1671(1990);
J. Isenberg and V. Moncrief, {\it Ann. of Phys.} (NY) {\bf 199}, 84(1990);
P. T. Chru\'sciel, On uniqueness in the large of
solutions of Einstein equations, Canberra: ANU Press 1992;
A. D. Rendall, {\it Ann. Phys.} {\bf 233}, 82(1994);
\bibitem{BMOM1988} P. Bizo\'n, E. Malec and N. \'O Murchadha, Phys. Rev. Lett.
{\bf 61} 1147(1988); Class. Quantum Grav. {\bf 6}, 961 (1989);
{\bf 7}, 1953(1990);
\bibitem{MOM1994b} E. Malec and N. \'O Murchadha,
{\it Phys. Rev. } {\bf 50}, R6033(1994);
\bibitem{MOM1994a} that is the content of the  Penrose inequality,
which has been proven in the spherically symmteric case, cf.
E. Malec and N. \'O Murchadha, Phys. Rev. {\bf D49}, 6931
(1994); see also a proof by S. Hayward (in Tolman coordinates), {\it Class.
Quantum Grav.} {\bf 11}, 3037(1994);
\bibitem{Christodoulou} D. Christodoulou, {\it Commun. Math. Phys.} {\bf 105},
337(1986); {\bf 106}, 587(1987); {\bf 109}, 613(1987);
\bibitem{Alan9295} G. Rein, A. D. Rendall and J. Schaeffer,
 {\it Commun. Math. Phys.} {\bf 168}, 467(1995);
G. Rein and  A. D. Rendall, {\it Commun. Math. Phys.} {\bf 150}, 561(1992);
\bibitem{MEnew} E. Malec, {\it  Self-gravitating
spherically symmetric scalar fields} (1996); 
\bibitem{Choquet-Bruhat} a local existence proof was firstly obtained
by Choquet-Bruhat, using the harmonic gauge, cf. Choquet-Bruhat in
Gravitation: An Introduction to Current Research, ed. L. Witten,
 New York: Wiley 1962.
\end{references}
\end{document}